\documentclass[12pt]{iopart}

\usepackage{graphicx}

\newcommand{\lesim}{\mbox{\raisebox{-0.6ex}{$\stackrel{<}{\sim}$}}\:}

\begin{document}

\title[Hydrodynamic models]{Hydrodynamic models}

\author{Tetsufumi Hirano
}

\address{RIKEN BNL Research Center, Brookhaven National Laboratory,
Upton, NY 11973, USA
}


\begin{abstract}
Recent developments 
based on relativistic hydrodynamic
models in high energy
heavy ion collisions are discussed.
I focus especially on
how hydrodynamics works at
RHIC energies and how one can use the most of it
in analyses of jet quenching and thermal electromagnetic radiations.
I also comment on improvement of initial conditions and viscosity
in hydrodynamic models.

\end{abstract}




\section{Introduction}

First data reported by the STAR Collaboration at 
RHIC \cite{STAR:v2} has a significant meaning that
the observed large magnitude of elliptic flow
for charged hadrons
is consistent with hydrodynamic predictions
\cite{KHHH}.
This suggests that
large pressure possibly in the partonic phase
is built at the early stage
($\tau \sim 0.6$ fm/$c$)
in Au+Au collisions at $\sqrt{s_{NN}} =$ 130 and 200 GeV.
This situation at RHIC is in contrast to
that at lower energies such as AGS or SPS
where hydrodynamics always overpredicts
the data.
Hadronic transport models
are very good to describe experimental data
at lower energies, while they fail to reproduce
such large values of
elliptic flow parameter at RHIC (see, e.g., Ref.~\cite{BS}).
So the importance of hydrodynamics is rising
in heavy ion physics.
In this review, I discuss first how well hydrodynamics
can fit the data in Sec.~\ref{sec:2}
 and next how hydrodynamics is related
with other non-flow phenomena such as jet quenching
and thermal photon emission in Sec.~\ref{sec:3}.
In Sec.~\ref{sec:4}, I argue several attempts to
incorporate initial conditions from other models.
Finally, I briefly mention about viscosity in Sec.~\ref{sec:5}.

\section{Results from the hydrodynamic approach at RHIC}
\label{sec:2}

After the first STAR data were published \cite{STAR:v2}, 
other groups at RHIC have also obtained the data
concerning with flow phenomena \cite{Retiere}.
To understand
these experimental data,
hydrodynamic analyses are also performed extensively \cite{QGP3:hydro}.
Here I pick up several
results on elliptic flow and emphasize on
the momentum regions where hydrodynamics
gives a good description. 
Elliptic flow is very sensitive to the degree of secondary
interactions \cite{O}. The indicator
is the second harmonic coefficient
of azimuthal distributions \cite{PV}
\begin{eqnarray}
v_2(p_T, y) & = & \frac{\int d\phi \cos(2\phi)\frac{dN}{p_Tdp_T dy d\phi}}{\int d\phi \frac{dN}{p_T dp_T dy d\phi}} = \langle \cos(2\phi) \rangle.
\label{eq:v2}
\end{eqnarray}
(For recent progress of higher harmonics, see Refs.~\cite{P,AP}.)
In ideal hydrodynamics, the mean free path of particles is assumed to
be zero. So the hydrodynamic prediction of $v_2$
should be maximum among transport theories.
The scaled elliptic flow,
which is defined as $v_2$ divided by
initial spatial eccentricity
$\varepsilon = \langle y^2-x^2\rangle/\langle x^2+y^2\rangle$,
becomes almost constant around 0.2 from hydrodynamic simulations \cite{KSH}.
Interestingly, the experimental data reach
this hydrodynamic limit for the first time
in central and semi-central collisions at RHIC energies~\cite{STAR:scaledv2}.
The data of scaled $v_2$ at various collision energies
monotonically increase with multiplicity per unit transverse area
$(1/S)dN/dy$ and, eventually, comes to the hydrodynamic limit.
The key quantity to achieve thermalisation
is the number of particles in a unit volume.
This can be seen also in the centrality dependence of $v_2$.
Deviation of the hydrodynamic result
from experimental data starts from
$N_{\mathrm{ch}}/N_{\mathrm{max}}\sim 0.5$
which corresponds to an impact parameter $b \sim 5$ fm \cite{KHHH}.
The $p_T$ dependence of $v_2$ contains
rich physics. In low $p_T$ region, $v_2$ increases with $p_T$
almost linearly.
On the other hand, the data points of 
$v_2$ saturate at high $p_T$ \cite{STAR:highptv2}
and deviate from hydrodynamic predictions.
In the intermediate $p_T$ regions, the interplay
between soft physics and hard physics
is very important in understanding
$v_2$ for identified hadrons.
This will be discussed in the next section.
Hydrodynamic predictions give good agreements
with experimental data of $v_2$ including mass dependences
in low $p_T$ region \cite{PHENIX:v2id}.
In low $p_T$ region at midrapidity, hydrodynamics
reproduces experimental data very well.
While, in forward/backward rapidity regions where
multiplicity becomes small,
hydrodynamics overpredicts largely \cite{Hirano}
the experimental data observed by
the PHOBOS Collaboration \cite{PHOBOS:v2eta}.
This suggests thermalisation is achieved only near
midrapidity.
The regions where hydrodynamics works well
for charged hadrons are summarised as follows:
$\mid b \mid \lesim 5$ fm, $p_T \lesim 1.5$ GeV and
$\mid \eta \mid \lesim 2$ \cite{QGP3:hydro}.
The scaled $v_2$ suggests that thermalisation is only partially achieved
in small $(1/S)dN/dy$.
Heinz tried to discuss a possible mechanism of
deviation between
hydrodynamic calculations and experimental data
in forward/backward rapidity regions or in peripheral collisions
by introducing ``thermalisation coefficient" \cite{Heinzqm2004}.
%

\section{Information inside fluids}
\label{sec:3}

Due to strong interaction among secondary
particles, hadrons can be emitted
only from freezeout hypersurface. 
So hadron spectra reflect the information
simply about accumulation of the space-time evolution.
Unlike the blast-wave model fitting,
the space-time
dependences of thermodynamic variables
are obtained in hydrodynamic simulations. 
These informations are significantly helpful
to understand what happens \textit{inside}
the reaction regions.
Here I discuss two phenomena, i.e.,
jet quenching and electromagnetic radiations,
which are directly related with information
inside bulk matter.

\subsection{Jet quenching}

Minijets produced in initial semihard collisions
have to traverse the reaction region
where bulk matter evolves.
During traveling, these minijets lose their energies
through interactions with the medium.
So high $p_T$ hadrons originated from fragmentation of minijets
contain information about the medium \cite{QGP3:highpt}.
Gyulassy \textit{et al.} first employed
results from (2+1)-dimensional
hydrodynamic simulations
in a jet quenching analysis \cite{GVWH}.
The first order term of an energy loss formula
in the opacity expansion becomes \cite{GLV}
\begin{equation}
\Delta E = C \int_{\tau_0}^{\infty} d\tau
\rho\left(\tau, \vec{x} \left(\tau\right)\right)
(\tau-\tau_0)\ln\left({\frac{2E_0}{\mu^2 L}}\right).
\label{eq:GLV}
\end{equation}
Here kinematics of emitted gluons are neglected.
A dimensionless parameter $C$ includes
strong running coupling constant
and colour Casimir factors.
One needs the space-time evolution
of parton density $\rho$ in quantitative analysis for
the energy loss of a parton.
Note that the parton density appeared in Eq.~(\ref{eq:GLV})
does not need to be thermalised because it
simply comes from the inverse mean free path of
an energetic parton.
Nevertheless,
the parton density in Eq.~(\ref{eq:GLV})
should be a thermalised one at RHIC energies
from the analyses of $v_2$ discussed in the previous section.
In Ref.~\cite{HiranoNara}, it is assumed that the parton density
in an energy loss formula
is a solution of hydrodynamic equations 
in full 3D space which is
compatible with low $p_T$ data
such as $dN/dy$ and $p_T$ spectra \cite{HiranoTsuda}
(the hydro+jet model).
Systematic studies
based on the hydro+jet model with Eq.~(\ref{eq:GLV})
are performed for
$p_T$ spectra \cite{HiranoNara5},
back-to-back correlation functions \cite{HiranoNara3},
pseudorapidity dependence
of nuclear modification factors \cite{HiranoNara4},
and $v_2$ \cite{HiranoNara5,HiranoNara4} in high $p_T$ regions.
It is not obvious where hydrodynamic $p_T$ spectrum
eventually turns into pQCD power law spectrum.
Moreover, the transition point can depend on particle
species. Interplay between radial flow and jet quenching
is discussed in Ref.~\cite{HiranoNara5}.
It is found that the transition point increases with
hadron mass due to mass dependent effects of radial flow.
\begin{figure}
\begin{center}
\includegraphics[width=0.47\textwidth]{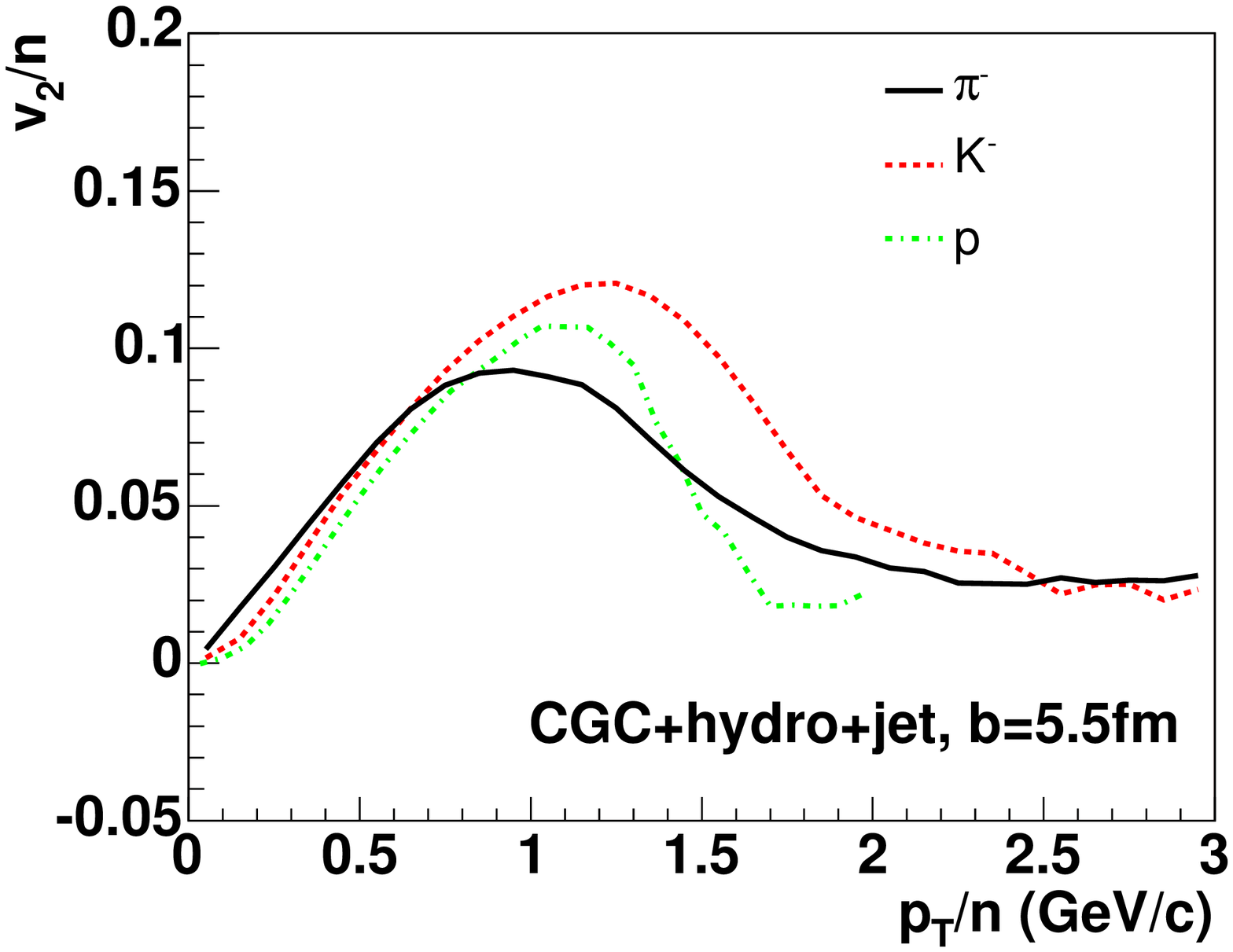}
\includegraphics[width=0.42\textwidth]{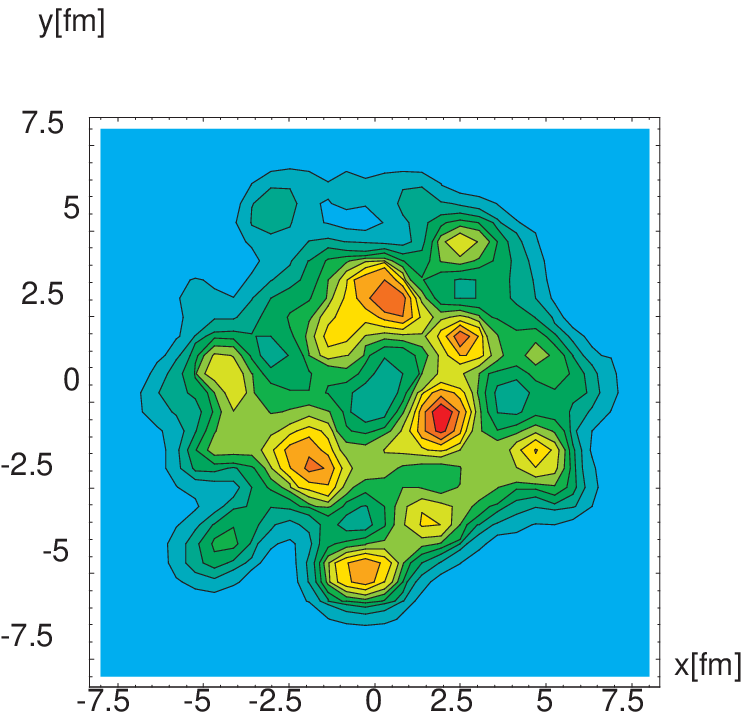}
\caption{
(Left) $v_2$ divided by
the number of constituent quark from the hydro+jet model.
 (Right) Initial transverse energy density distribution
 of one specific event
 in a central Au+Au collision
 at $\sqrt{s_{NN}}=200$ GeV from NeXus.
 Courtesy of SPheRIO collaboration.}
\end{center}
\end{figure}
The resultant $v_2$ divided by the number of constituent quarks
is represented in Fig.~1 (left).
This looks very similar to the results from
recombination/coalescence models \cite{RF}.
The nuclear modification factor $R_{CP}$
for $\phi$ mesons is 
found to be 
smaller than for protons even though $m_{\phi}>m_{p}$ \cite{JV}.
This may be the meson-baryon effect \cite{RF} rather than
the particle mass effect resulting from radial flow.
But there is a possibility that
$\phi$ mesons do not participate in the hydrodynamic flow.
This is suggested by a hybrid model in which
hadronic afterburner is described by RQMD \cite{TLS}.
Theoretically, one needs to study hadronic afterburner
by combining a hadronic cascade model 
with the hydro+jet approach.
Experimentally, it is important to observe $v_2$ for $\phi$
mesons
in very low $p_T$ region where hydrodynamics is expected to work.
These will reveal the production
mechanism in the intermediate $p_T$ region at RHIC.

\subsection{Electromagnetic radiation}

Photons and dileptons emitted from thermalised matter
are signals of the QGP in heavy ion physics
proposed many years ago \cite{Shuryak}.
Here I focus on the discussion of thermal photon emissions.
As emphasised by Moore \cite{GM},
the yield of thermal photons is much more sensitive
to temperature profile and space-time volume
than the production rate 
(the invariant yield per unit space-time volume).
In a hydrodynamic approach, the invariant yield of thermal photon
is evaluated by accumulation of the production rate
over all fluid elements above
the kinetic freezeout temperature
\begin{eqnarray}
E\frac{dN_\gamma}{d^3p} = \sum_i \Delta V_i 
\tilde{E}\frac{dR_\gamma}{d^3\tilde{p}}(T(x_i),\mu(x_i),
\tilde{E} =p_\mu u^\mu(x_i))
\end{eqnarray}
Here $\tilde{E}$ and $\tilde{p}$ are, respectively, 
energy and momentum of
an emitted photon measured in a local rest frame.
Information from hydrodynamic simulations
is taken through
temperature $T$,
chemical potential $\mu$ (or fugacity $\lambda$),
four fluid velocity $u^\mu$
and four dimensional volume $\Delta V_i$
for each fluid element at $x_i$.
A novel calculation of thermal photon yield is based on
the hydrodynamics with rate equations
for quarks and gluons \cite{biro,Niemi}.
At collider energies, the QGP at the very early stage is
supposed to be the ``gluon plasma (GP)"
since
multiparticle production
is dominated by small $x$ gluons
in the wave function of colliding nuclei
at ultrarelativistic energies.
Initial gluons achieve a thermalised state,
but it may not be a chemically equilibrium one yet.
Through the process $gg \rightarrow q\bar{q}$,
the system of the GP goes towards to the QGP.
Due to the smaller degree of freedom,
the initial temperature in the GP is larger than
in the QGP when one assumes a fixed initial energy density.
The photon yield would be enhanced due to
higher initial temperature.
On the other hand, small values of 
fugacity $\lambda$ for quarks and gluons
make the photon yield per unit space-time volume
suppressed.
The competition between these two effects
is studied in terms of a hydrodynamic model
with rate equations.
It is found that unexpected cancellation happens
between initial temperature effect and fugacity effect 
in the final spectrum \cite{Niemi}.
I note that it is worth studying thermal photon yields from
chemically non-equilibrium \textit{hadronic} phase
in order to conclude
real excess of thermal photon
yield from the QGP phase.
Related with chemical non-equilibrium properties,
it is interesting to see
whether hydrodynamic results for elliptic flow
will be changed at RHIC due to the insufficient chemical
equilibrium.

\section{Improvement of initial conditions in hydrodynamic models}
\label{sec:4}

All the hydrodynamic
results mentioned in Sec.~\ref{sec:2} are based on
ideal hydrodynamics 
with parametrised initial conditions which
lead to the reproduction of the multiplicity and
$p_T$ spectra for hadrons
in low $p_T$ region.
Towards description of heavy ion 
collisions from initial colliding nuclei
to final spectra in a unified way,
it is much better to
employ effective theories/models
which are relevant in the very early stage
of collisions.
There are some attempts to incorporate
results from event generators, e.g., HIJING \cite{GRZ},
URASiMA \cite{NHM}, VNI \cite{SS} or NeXus \cite{OAHK},
 into initial conditions of
hydrodynamic simulations.
In other approaches, effective theories,
e.g., string ropes/flux tubes \cite{MCS},
the final state saturation \cite{ERRT}
or the initial state saturation \cite{Nara}, are employed. 
Here I pick up two approaches which are, to my mind, important
for the RHIC physics.

\subsection{Fluctuation of initial conditions}

In conventional hydrodynamic simulations for non-central collisions,
one parametrises
initial transverse profile by using
the number density of participants or binary collisions
at a fixed impact parameter.
This gives us a smooth profile for thermodynamic
variables.
However, if one picks up one event, the energy density distribution
in the transverse plane has an extremely bumpy
structure
shown in Fig.~1 (right).
In SPheRIO \cite{OAHK}, initial conditions for hydrodynamic
simulations are taken from an event generator
NeXus \cite{NeXus} as an event-by-event basis.
The final spectra in this approach are
calculated like conventional event generators:
they perform hydrodynamic simulations,
calculate particle spectra at each event, accumulate many events,
and average particle spectra over simulated events.
It is worth noting that the numerical cost
of the hydrodynamic code 
SPheRIO is very cheap
and that they are able to simulate
many events even in full 3D space.
It is found that the multiplicity
from initial energy densities with fluctuations
are always smaller than without fluctuations.
One can easily show \cite{Hama} the following relation
between initial and final total entropy
in a simple EoS case:
$\langle S \rangle_\mathrm{final}  \sim 
\langle S \rangle_\mathrm{initial}
\left[1- \alpha\frac{\langle \Delta E^2\rangle}
{\langle E \rangle^2}\right]$
where $\langle \cdots \rangle$ means the event average
and $\alpha$ is a positive constant.
One should keep this fact in mind in detailed analyses
based on hydrodynamic models.
Needless to say, a systematic study of 
the effect of fluctuation on $v_2$
would be interesting since elliptic flow is sensitive to
the initial geometry.

\subsection{Initial conditions from the colour glass condensate}

Jet quenching and large elliptic flow are
two important findings measured at RHIC.
The common key for these two phenomena
is the dense partonic medium.
What is a possible origin of this dense matter
in Au+Au collisions at RHIC?
This can be traced to the initial parton
density inside the colliding nuclei.
It is well known that small $x$ gluons
are dominant for multi-particle production.
In the ultra-relativistic limit,
these gluons in colliding nuclei could
form the Colour Glass Condensate (CGC) \cite{MV,Jamal}.
So initial conditions in relativistic heavy ion collisions
can be described by melting the CGC \cite{QGP3:cgc}.
Hirano and Nara employed
a nuclear wave function discussed in Ref.~\cite{KL}
and calculate gluon number density
at each transverse point
through $k_T$ factorised formula.
Regarding this gluon number density as a thermalised one,
they evaluate energy density at each space point
at initial time $\tau_0$.
By throwing it into hydrodynamic simulations
as an initial condition, they obtain 
pseudorapidity distribution
in the usual hydrodynamic manner \cite{Nara} 
and compare these results
with the PHOBOS data \cite{PHOBOS:dndeta}.
The CGC initial conditions
are found to work very well for describing
the energy, centrality, and (pseudo)rapidity
dependences of charged hadrons \cite{Nara}.

\section{Viscosity}
\label{sec:5}

Finally, I briefly mention about the viscosity.
There is no a priori reason why ideal hydrodynamics
works so well at RHIC as discussed in Sec.~\ref{sec:2}.
(For a perspective from a strong coupling
gauge theory, see Refs.~\cite{PSS,S}.)
Viscous corrections to
momentum distribution functions
are considered in Refs.~\cite{A,T}.
The first order correction to the momentum distribution function
becomes 
$\delta f \propto \frac{\Gamma_s}{T^2}f_0(1+f_0)p^\mu p^\nu X_{\mu\nu}$,
where $\Gamma_s = \frac{4}{3} \eta/(e+p)$
is the sound attenuation length, $f_0$ is the Bose
distribution function, and $X_{\mu\nu}$ is the tensor part of
the thermodynamic force.
Relation between the inverse Reynolds number \cite{Baym}
and the attenuation length becomes
$R^{-1} \approx \Gamma_s/\tau$ in the Bjorken flow case \cite{KMTS}.
$R^{-1}$ is found to be very small from the blast-wave
fitting with the above correction term \cite{T}.
This suggests that the hadronic fluid in heavy ion collisions
is nearly perfect one.
It is known that naive relativistic extension of
Navier-Stokes equations breaks down due to infinite
signal velocity since the equations are parabolic ones
(see, e.g., Ref.~\cite{Strottman}).
One can introduce relaxation terms to avoid this problem \cite{T2}.
The existence of relaxation terms are essential
since they change the type of equations to hyperbolic
ones in which signal velocity remains finite.
More systematically,
one takes account of the second order viscous
terms \cite{M} 
from the prescription of 
extended thermodynamics (see, e.g., Ref.~\cite{Muller}).
Dynamical studies of viscous fluids are mandatory for
comprehensive understanding of the QGP.

\section{Summary and Discussion}

I reviewed the hydrodynamic results
at RHIC and discussed some related topics such as
jet quenching and electromagnetic radiations
from a hydrodynamic
point of view.
I also discussed recent attempts for initial conditions
of hydrodynamic simulations
and for viscous effects. 
In addition to the photon emission discussed in Sec.~\ref{sec:3},
dileptons are also interesting to study chiral properties
of hadrons by using
hydrodynamics. In the low invariant mass regions, dilepton
spectrum is the best tool to see the spectral change of
hadrons in hot/dense medium (see, e.g., Ref.~\cite{Hatsuda}).
On the other hand,
in the high invariant mass region,
it is interesting to see whether $J/\psi$ really melts
in the QGP phase \cite{MS,AH,K}. Hydrodynamics
provides the temperature profile
and can be also useful for quantitative analyses
of these phenomena.
Some open questions are as follows:
Usually, the initial time
in hydrodynamic simulations
is chosen as around $\tau_0\sim 0.6$-$1.0$ fm/$c$.
How do we get such an early thermalisation time?
Although $gg \rightarrow ggg$ process
seems to play
an important role in thermalisation,
the resultant thermalisation time
is still a few fm/$c$ \cite{bottomup}.
Note that a typical life time of the QGP phase
from hydrodynamic simulations is around
3 fm/$c$ in central collisions at RHIC.
In this review, I did not go into details
about the HBT puzzle \cite{DM}.
The important point is to find a solution of the HBT puzzle
which is compatible with
other observables such as $p_T$ spectra and $v_2$.

\ack
I am grateful to Y.~Hama, U.~Heinz, T.~Kodama, 
Y.~Nara, H.~Niemi and D.~Teaney for fruitful discussions
to prepare my review talk.
Special thanks go to L.~McLerran and Y.~Nara
for continuous encouragement.
This work is supported by RIKEN.

\end{document}